\documentclass{raa}           
\usepackage{graphicx,times}
\usepackage{natbib}
\usepackage{amssymb,amsmath}
\bibpunct{(}{)}{;}{a}{}{,}

\usepackage[a4paper=true,dvipdfm=true,pagebackref=true]{hyperref}
\hypersetup{pdftitle = Post-SN RSC, pdfauthor = Y. Zeng, pdfsubject= The subject, pdfkeywords = keyword1 keyword2 keyword3} 
\hypersetup{colorlinks = true, linkcolor = orange, anchorcolor = red, citecolor = blue, filecolor = red, pagecolor = red, urlcolor = black}

\usepackage[utf8]{inputenc}

\begin{document}

   \title{On the rotation properties of a post-explosion helium-star companion in Type Iax supernovae}

 \volnopage{ {\bf 20XX} Vol.\ {\bf X} No. {\bf XX}, 000--000}
   \setcounter{page}{1}

   \author{Yaotian Zeng\inst{1,2,3}, Zheng-Wei Liu\inst{1,2,4}, Xiangcun Meng\inst{1,2,4}, Zhanwen Han\inst{1,2,3,4}}

   \institute{Yunnan Observatories, Chinese Academy of Sciences (CAS), Kunming 650216, P.R.~China; {\it zengyaotian@ynao.ac.cn; zwliu@ynao.ac.cn}\\
             \and
             Key Laboratory for the Structure and Evolution of Celestial Objects, CAS, Kunming 650216, P.R.~China\\
	         \and
             University of Chinese Academy of Science, Beijing 100012, P.R.~China\\
             \and 
             Center for Astronomical Mega-Science, CAS, Beijing 100012, P.R.~China\\
\vs \no
   {\small }
}

\abstract{Recent studies have suggested that type Iax supernovae (SNe Iax) are likely to result from a weak deflagration explosion of a Chandrasekhar-mass white dwarf in a binary system with a helium (He)-star companion. Assuming that most SNe Iax are produced from this scenario, in this work we extend our previous work on the three-dimensional hydrodynamical simulation of ejecta-companion interaction by taking the orbital and spin velocities of the progenitor system into account. We then follow the post-impact evolution of a surviving He-star companion by using the one-dimensional stellar evolution code \textsc{MESA}. We aim to investigate the post-explosion rotation properties of a He-star companion in SNe Iax. We find that the He-star companion spins down after the impact due to the angular-momentum loss and expansion caused by the mass-stripping and shock heating during the interaction. This leads to the situation where the surface rotational speed of the surviving companion can drop to one-third of its pre-explosion value when it expands to a maximum radius a few years after the impact. Subsequently, the star shrinks and spins up again once the deposited energy is released. This spin-switching feature of the surviving He-star companions of SNe Iax may be useful for the identification of such objects in future observations. 
}

   \authorrunning{Y. Zeng et al.}            
   \titlerunning{Rotation of a surviving He-star companion}  
   \maketitle

%
\section{Introduction}           
\label{sect:intro}

Most type Ia supernovae (SNe Ia) follow an empirical width-luminosity relation, i.e., the so-called `Phillips relation' \citep{Phillips1993ApJ}. These SNe Ia are usually called as `normal' SNe Ia. Using normal SNe Ia as a good cosmic distance indicator has led to the discovery of the accelerating expansion of the Universe \citep{Riess1998AJ, Schmidt1998ApJ, Perlmutter1999ApJ}.  However, the progenitor models and explosion mechanism of SNe Ia are still unknown. An SN Ia is generally thought to be a thermonuclear explosion of a near-Chandrasekhar-mass or sub-Chandrasekhar-mass white dwarf (WD) through accreting material either from a non-degenerate companion star (i.e., the single-degenerate [SD] model; \citealt{Whelan1973ApJ, Han2004MNRAS,Liu2018MNRAS,Liu2020AA}) or another WD (i.e., the double-degenerate model; \citealt{Iben1984ApJS,Webbink1984ApJ}). In the SD model, the non-degenerate companion star could be a main sequence star, a red giant star or a helium (He) star, and the companion star is expected to survive from the explosion \citep[e.g.,][]{Wheeler1975ApJ,Marietta2000ApJS, Podsiadlowski2003,Han2008,Pakmor2008AAP,Liu2012AAP, Liu2013ApJa,Liu2013AAP, Liu2013ApJb,Liu2021b,Pan2012ApJ,Shappee2013ApJ,Boehner2017MNRAS,Bauer2019ApJ}.

More and more observations have shown that there are sub-classes of SNe Ia \citep[e.g.,][]{Filippenko1992AJ, Filippenko1992ApJ, Li2003PASP}. Type Iax supernovae (SNe Iax) are the most common sub-class of SNe Ia \citep[][]{Foley2013ApJ,Liu2015MNRAS}. To date, about $50$ SNe Iax have been found \citep[e.g.,][]{Jha2017Hsn}, which contribute around $30\,\%$ of the total SN Ia birthrates \citep{Li2003PASP, Foley2013ApJ, White2015ApJ}. Recent studies seem to suggest that SNe Iax are produced from the weak deflagration explosion of a near-Chandrasekhar-mass WD in an SD binary system with a He-star companion. For instance, He emission lines have been detected in early-time spectra of two SNe Iax, i.e., SN~2004cs and SN~2007J \citep{Rajala2005PASP, Foley2009AJ, Foley2013ApJ}. A possible progenitor He-star companion has been detected in the pre-explosion images of an SN Iax SN~2012Z \citep{McCully2014Nature, Liu2015ApJ}. Most SNe Iax have been found in late-type, star-forming galaxies \citep[e.g.,][]{Foley2010AJ, Foley2013ApJ}, suggesting short delay times for SNe Iax that are consistent with the theoretical predictions of the progenitor systems composed of a WD and a He-star companion \citep[e.g.,][]{Foley2013ApJ, Lyman2013MNRAS, Lyman2018MNRAS, White2015ApJ,Takaro2020MNRAS,Liu2015AAP}.  In addition,  recent studies have shown that the weak deflagration explosion of a near-Chandrasekhar-mass WD seems to be able to well reproduce the observational features of SNe Iax \citep[e.g.,][]{Branch2004PASP,Jordan2012ApJL,Kromer2013MNRAS,Fink2014MNRAS}. However, other possible models such as the pulsational delayed detonation explosion model have also been suggested for SNe Iax \citep[][]{Hoeflich1995ApJ,Meng2014ApJL,Stritzinger2015AAP}.

In our previous work of \citet[][]{Zeng2020ApJ}, by assuming that SNe Iax are caused by weak deflagration explosions of progenitor systems composed of a WD and a He-star companion, we have investigated the details of ejecta-companion interaction by performing three-dimensional (3D) hydrodynamical simulations with the smoothed particle hydrodynamics \citep[SPH;][]{Gingold1977MNRAS, Lucy1977AJ} code \textsc{Stellar GADGET} \citep{Pakmor2012MNRAS}. We find that a small amount of He mass ($\sim0.4\%$ of companion masses) is stripped off from the companion surface during the ejecta-companion interaction \citep[][]{Zeng2020ApJ}, which provides an explanation for the non-detection of He lines cased by the swept-up He-rich companion material in late-time spectra of SNe Iax \citep[e.g.,][]{Foley2013ApJ, Foley2016MNRAS, Magee2019AAP, Jacobson2019MNRAS, Tucker2020MNRAS}. Furthermore, we have also followed the long-term evolution of the surviving He-star companions by using the one-dimensional (1D) stellar evolution code \textsc{MESA} \citep{Paxton2018ApJS} to predict their post-explosion properties \citep{ZengInprep}. However, in these previous studies we did not consider the orbital and spin velocities of the progenitor system in our 3D hydrodynamical impact simulations, leading to the situation that we could not completely study the post-impact rotation properties of a surviving He-star companion. It has been suggested that the rotation of a companion star can be significantly reduced due to the angular-momentum loss and significant expansion caused by the ejecta-companion interaction \citep[][]{Meng2011SCPMA,Liu2013AAP,Liu2017MNRAS,Liu2022ApJ,Pan2012ApJ,Pan2013ApJ}. For example, Tycho~G has been suggested to be a candidate of the surviving companion star in SN~1572 (i.e., Tycho's SN) because of its peculiar spatial velocity \citep{Ruiz-Lapuente2004Nature}. However, \citet{Kerzendorf2009ApJ} found that the Tycho~G has a small rotational velocity of $8\,\mathrm{km\,s^{-1}}$ \citep{Kerzendorf2014ApJ}, which is much slower than the predicted spin velocities of companion stars at the moment of SN explosion \citep{Han2008}. By investigating the post-impact rotation properties of the surviving companions, \citet{Pan2012ApJ} have suggested that one cannot rule out Tycho~G as the candidate of the surviving companion star of the Tycho SN Ia only based on its small rotational velocity because of the angular-momentum loss and significant expansion of the star due to the interaction can cause its rotational velocity drops significantly after the impact \citep{Pan2012ApJ, Liu2013AAP, Liu2022ApJ}. Furthermore, the post-impact rotation properties of an SN Ia's surviving companion star have been thought to be helpful for the identification of such objects in nearby SN remnants (SNRs).

In this work, by adopting the same progenitor and explosion models for SNe Iax that were used in \citet{Zeng2020ApJ}, we extend our previous 3D hydrodynamical simulations of ejecta-companion interaction by taking the orbital and spin velocities of the progenitor system into account. The main goal of this work is to investigate the post-impact rotation properties of a surviving He-star companion in SNe Iax. In Section~\ref{sec:method}, we briefly describe our methods and models.  The results of evolution of our surviving He-star companion model are given in Section~\ref{sect:results}, including evolutionary tracks, post-impact rotation properties and a comparison with the non-rotating model. Finally, we provide a summary and conclusion in Section~\ref{sect:summary}.

\section{Numerical methods}
\label{sec:method}

\begin{figure*}[t]
		\centering
		{\includegraphics[width=0.33\textwidth]{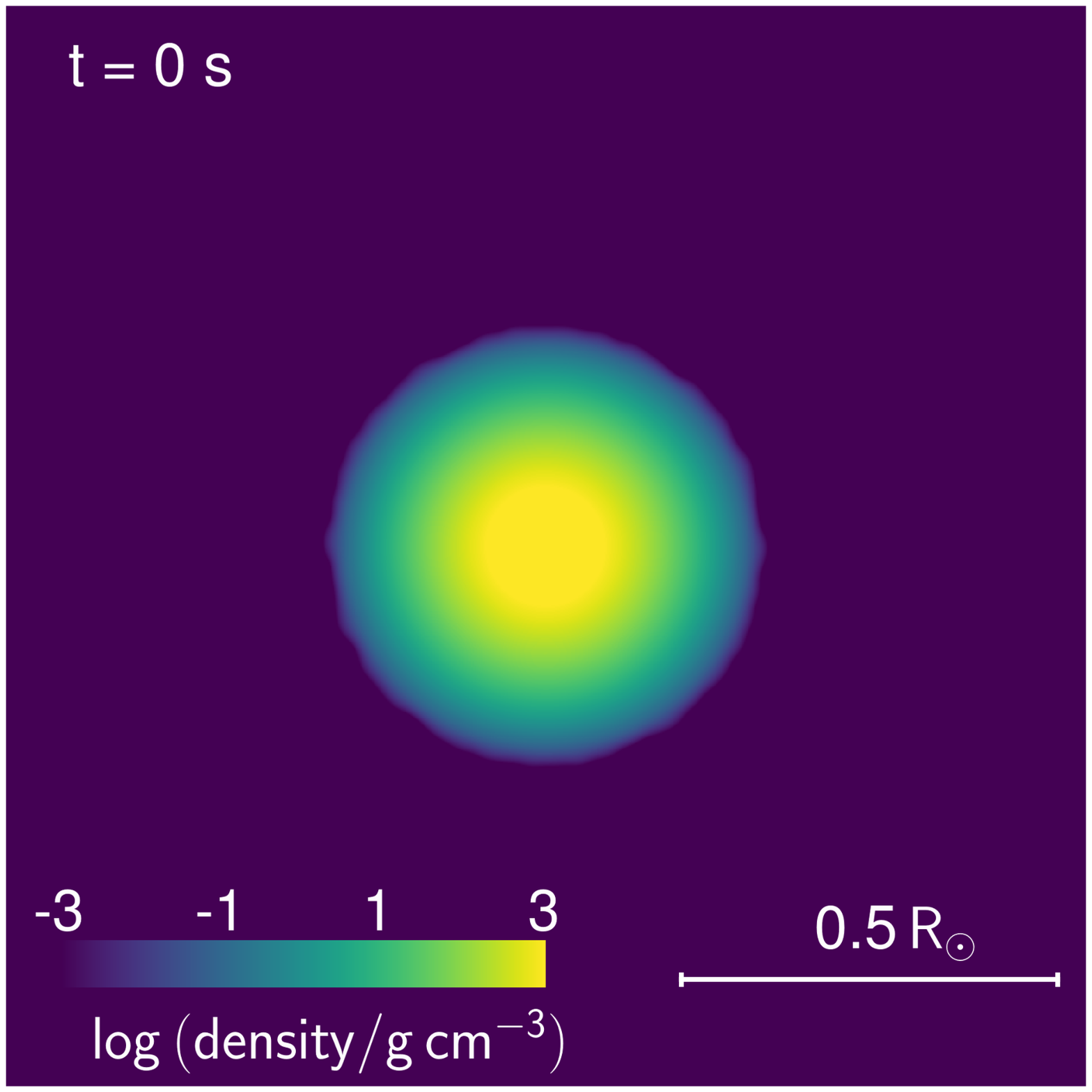}}
		{\includegraphics[width=0.33\textwidth]{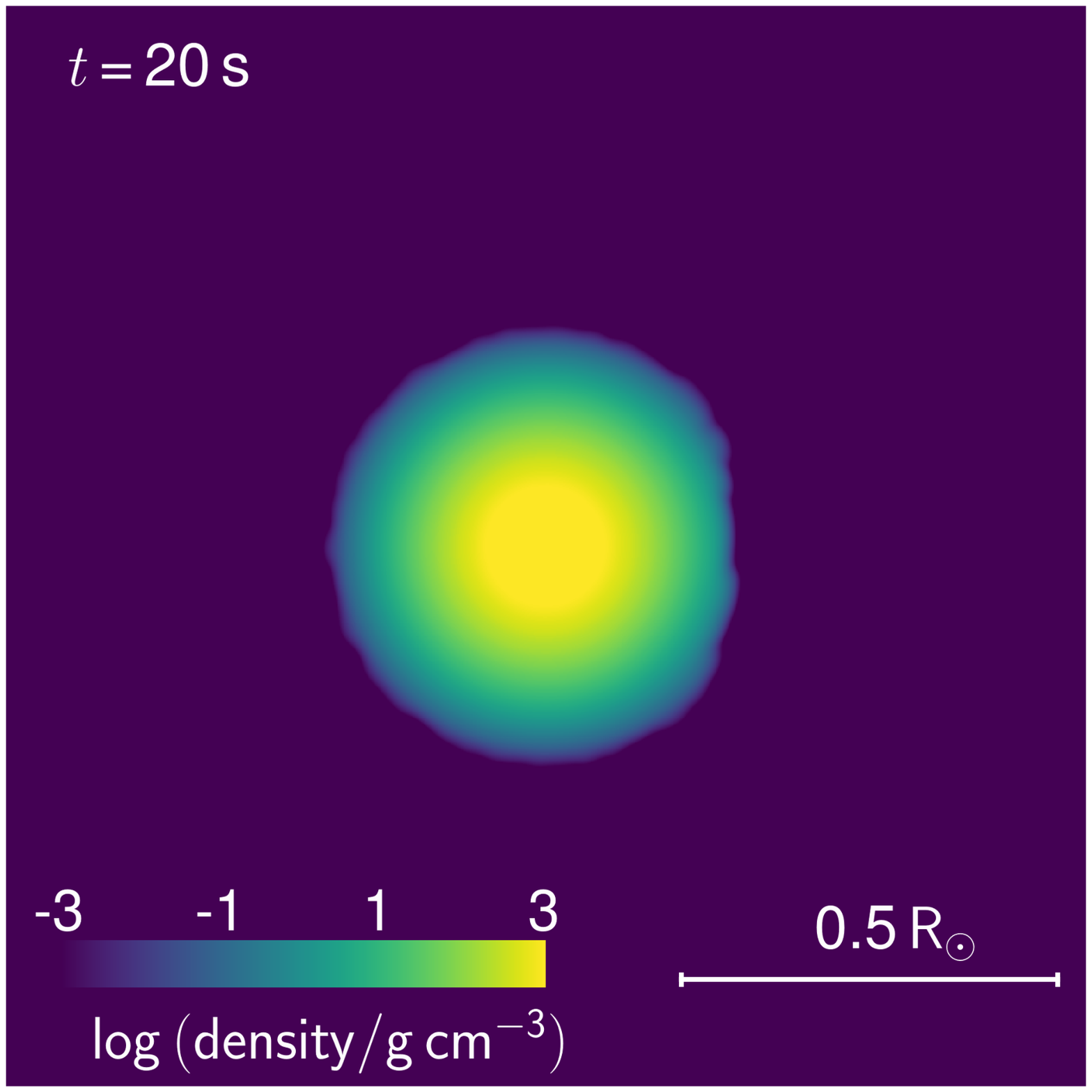}}
		{\includegraphics[width=0.33\textwidth]{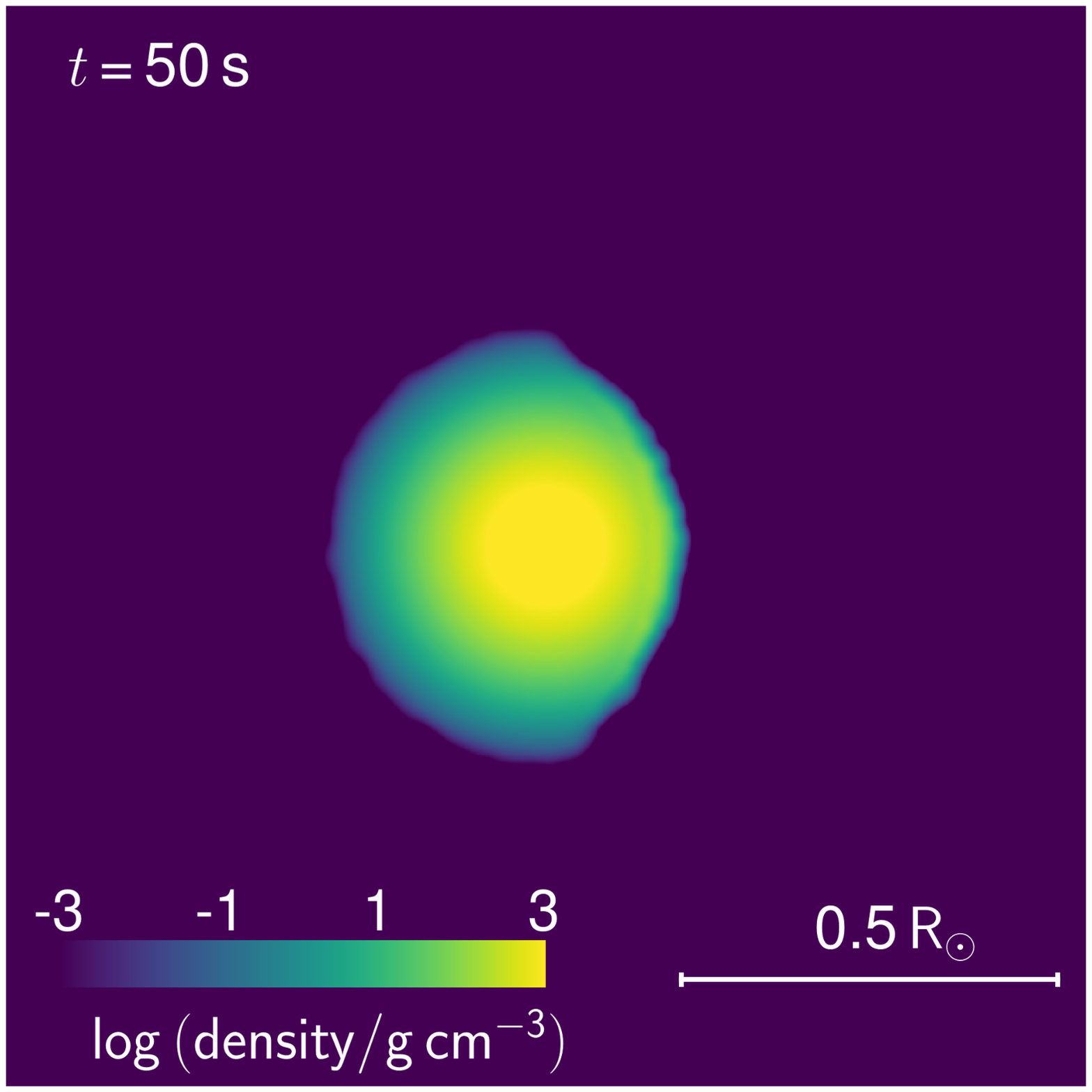}}
		{\includegraphics[width=0.33\textwidth]{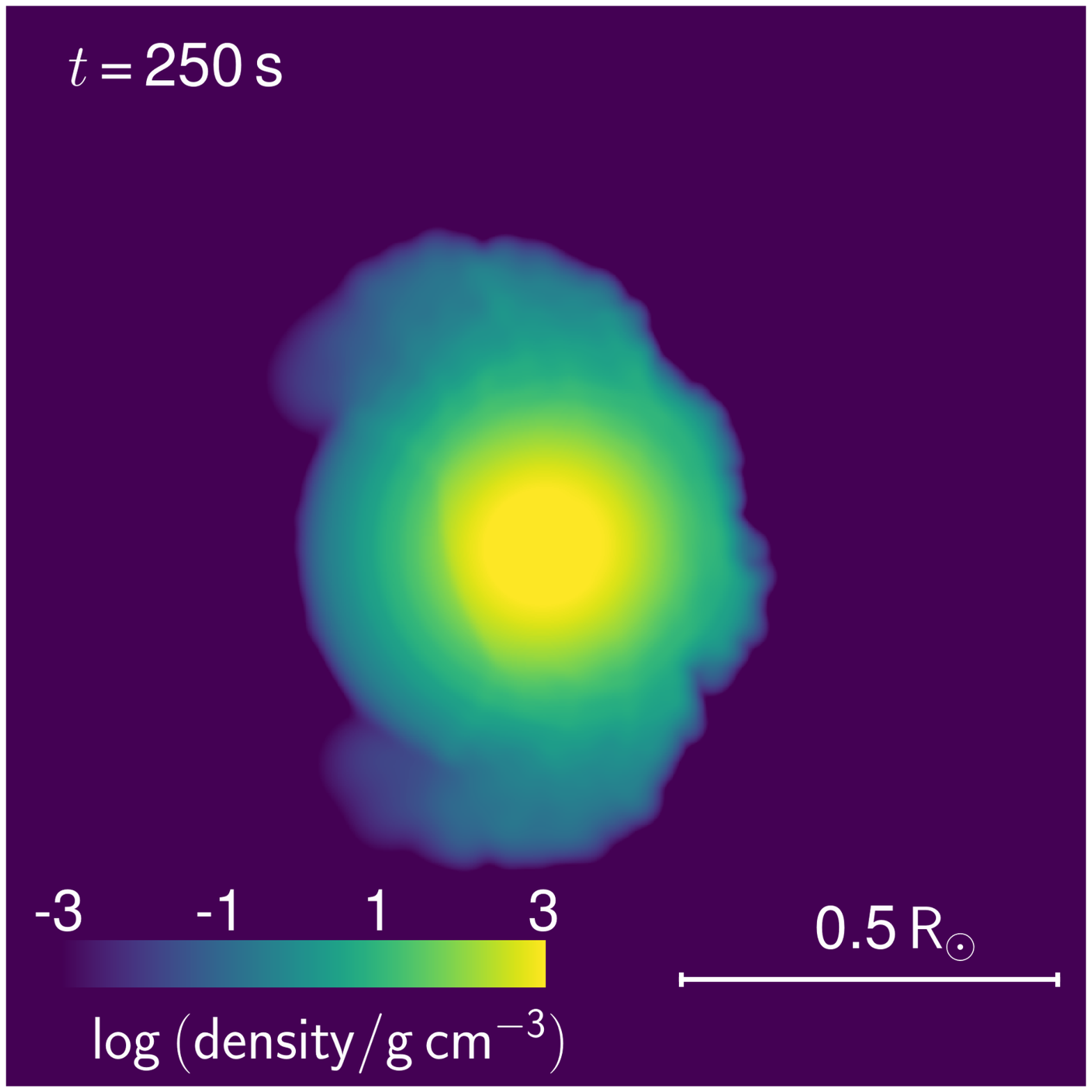}}
		{\includegraphics[width=0.33\textwidth]{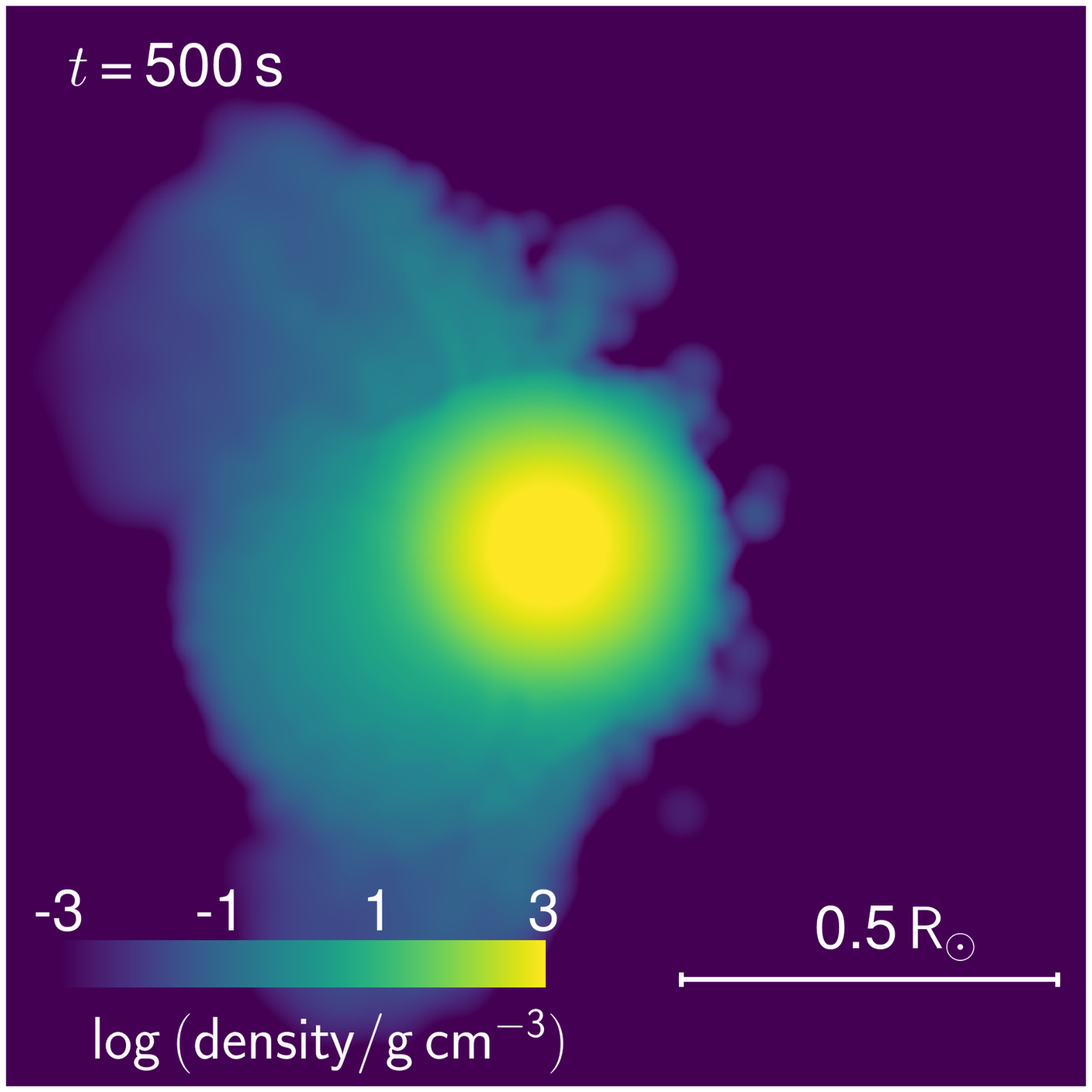}}
		{\includegraphics[width=0.33\textwidth]{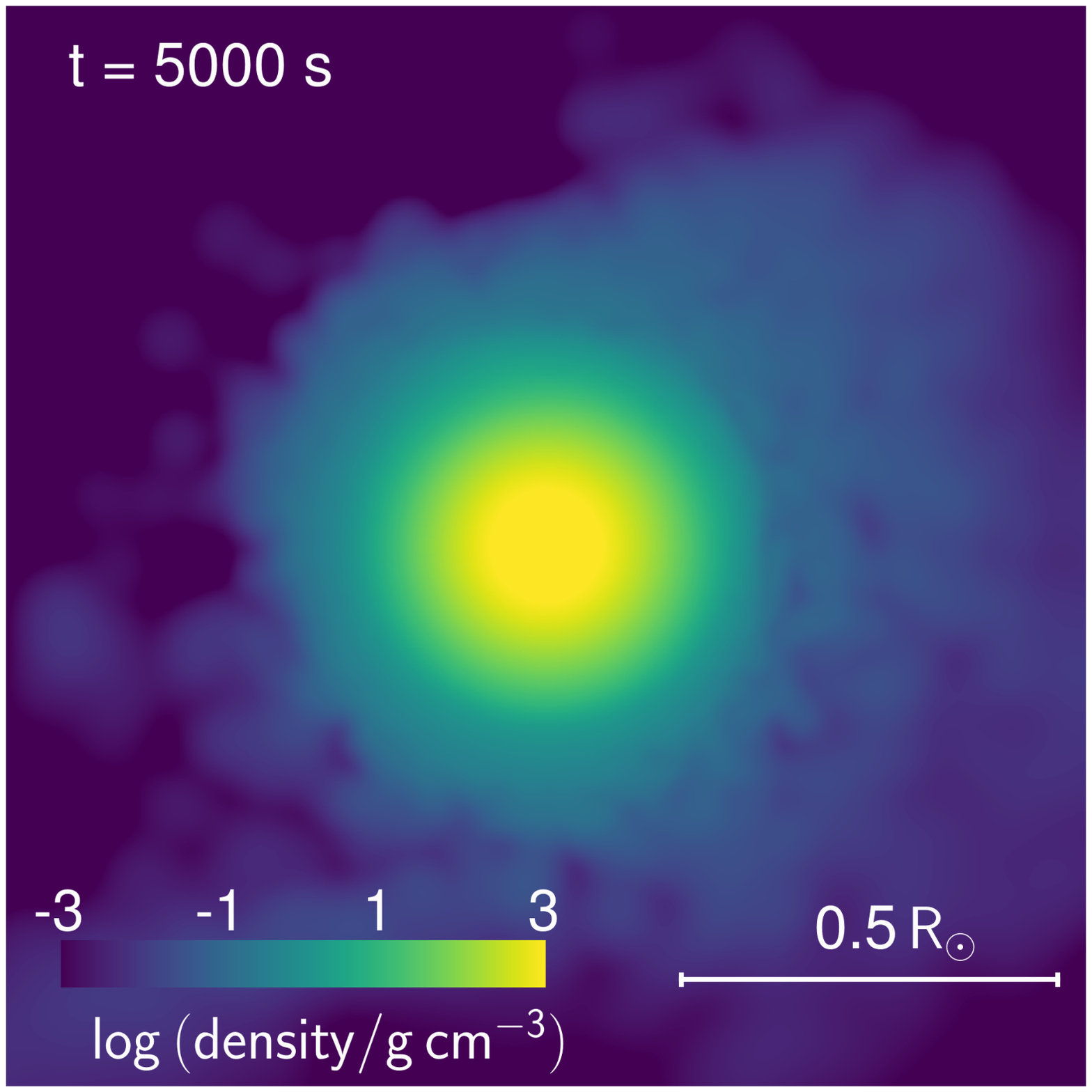}}
		\caption{\label{fig:1} Density distributions of our $1.24\,\mathrm{M}_{\odot}$ He-star companion model in the orbital plane at $t = 0\,\mathrm{s}$, $20\,\mathrm{s}$, $50\,\mathrm{s}$, $250\,\mathrm{s}$, $500\,\mathrm{s}$ and $5\mathord,000\,\mathrm{s}$ of our 3D SPH hydrodynamical impact simulation. The color scale indicates the logarithm of the mass density.}
\end{figure*}

\begin{figure}[t]
	\centering
	\includegraphics[width=1.0\textwidth]{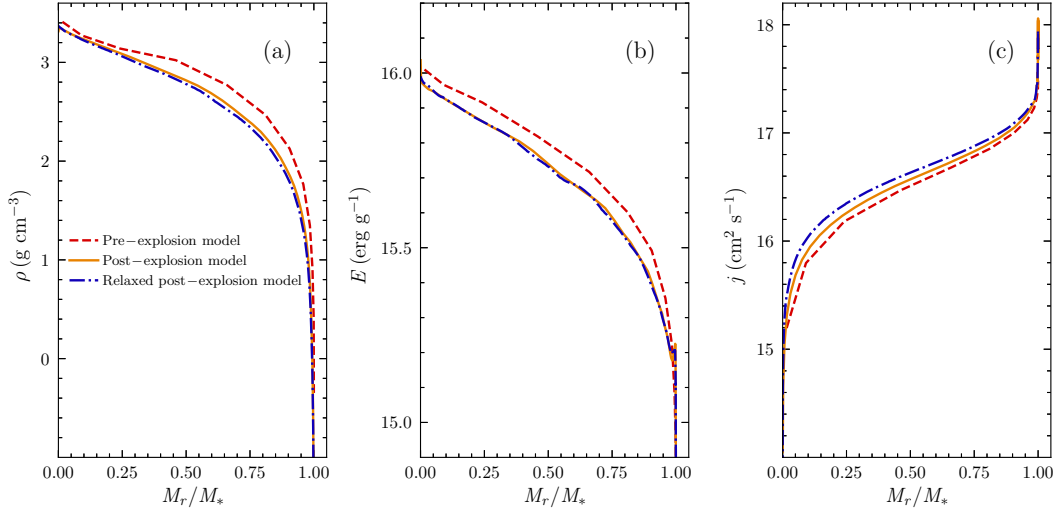}
	\caption{\label{fig:2} Radial profiles of density ($\rho$, left panel), specific internal energy ($E$, middle panel) and specific angular momentum ($j$, right panel) of our He-star companion model at the beginning (red dashed line) and the end of 3D SPH impact simulation (orange solid line). The corresponding profiles of the starting model for \textsc{MESA} post-explosion calculation are also drawn in blue dash-dotted lines. Here, $M_{\mathrm{r}}$ is the enclosed mass within a sphere of radius $r$, and $M_{\ast}$ is the total mass of the star.}
\end{figure}

\subsection{Ejecta-companion interaction}

To perform a 3D hydrodynamical simulation of ejecta-companion interaction, we employ the \textsc{Stellar GADGET} code \citep{Pakmor2012MNRAS}. The initial models and basic assumptions are set to be the same as those in our previous study \citep[for the details, see Section~2 of][]{Zeng2020ApJ} except for that the orbital and spin velocities of the progenitor system are considered in the present work. Therefore, we only briefly describe them as follows.

Our initial He-star companion model at the moment of SN Ia explosion was constructed by following 1D detailed binary evolution of a progenitor system composed of a $1.10\,\mathrm{M_{\odot}}$ WD and a $1.55\,\mathrm{M_{\odot}}$ He-star donor, in which the binary system has an initial orbital period of $\sim0.05\,\mathrm{day}$. The WD accretes He-rich material from the companion star through Roche-lobe overflow to increase its mass to $1.38\,\mathrm{M_{\odot}}$. At that point, we assume that the WD explodes as an SN Iax, and we take out the He-star companion model at this moment as the input of our subsequent 3D impact simulation \citep[for a detailed description, see][]{Liu2013ApJa}.  In addition, we assume that the rotation of the He-star companion is synchronized with its orbital motion at the time of SN Iax explosion due to strong tidal interaction during the pre-explosion mass-transfer phase, $\omega_{\rm rot} = \omega_{\rm orb}$, where $\omega_{\rm rot}$ and $\omega_{\rm orb}$ are the rotational and orbital angular velocities of the He-star companion, respectively. Based on our 1D full binary evolution calculation, the binary system at the moment of SN Ia explosion has a separation of $A$ = $\mathrm{5.16\times10^{10}\,cm}$, a WD mass of $M_{\rm WD} = \mathrm{1.38\,M_{\odot}}$, a companion mass of $M_{2} = \mathrm{1.24\,M_{\odot}}$, a companion radius of $\mathit{R}_{2}=1.91\times10^{10}\,\mathrm{cm}$, and a corresponding angular velocity of $\omega_{\rm rot}=1.59\times10^{-3}\,\mathrm{rad\ s^{-1}}$ (which leads to a companion rotational velocity of $\sim301\,\mathrm{km\ s^{-1}}$). We use the so-called `N5def model' to represent an SN Iax explosion in our impact simulations because this model has been found to well reproduce the observational features of a typical SN Iax such as SN~2005hk \citep{Kromer2013MNRAS,Fink2014MNRAS}. In the N5def model, the weak deflagration explosion does not completely disintegrate the entire Chandrasekhar-mass WD ( $\mathrm{1.40\,M_{\odot}}$), leaving a bound remnant WD of about $\mathrm{1.03\,M_{\odot}}$ after the explosion \citep{Kromer2013MNRAS,Fink2014MNRAS}.

We use the Cartesian coordinates in our 3D impact simulations. The $x$--$y$ plane is chosen as the orbital plane of the system, and the positive direction of the z-axis is the direction of the spin of the binary system. We add the initial orbital and spin velocities of each SPH particle by assuming that the binary system is synchronized. Therefore, the initial orbital and spin velocities of a particle $i$ are given by $\boldsymbol{v}_{i}=\boldsymbol{\omega}\times(\boldsymbol{r}_{i}-\boldsymbol{r}_{c})$, where $\boldsymbol{r}_{i}$ is the position of the particle $i$, $\boldsymbol{r}_{c}$ is the position of the center of mass of the binary system and $\omega$ is the angular velocity, $\omega=\omega_{\rm rot} = \omega_{\rm orb}$.  Similar to our previous work of \citet{Zeng2020ApJ}, in this work we also consider the mass-stripping due to the conservation of angular momentum and the shock heating and the energy deposition into the companion star in 3D impact simulations.  The 3D simulation is run for about $5\mathord,000\,\mathrm{s}$ after the explosion, at which time the unbound companion masses reach to a stable value.

\begin{figure*}[t]
	\centering
	\includegraphics[width=1.0\textwidth]{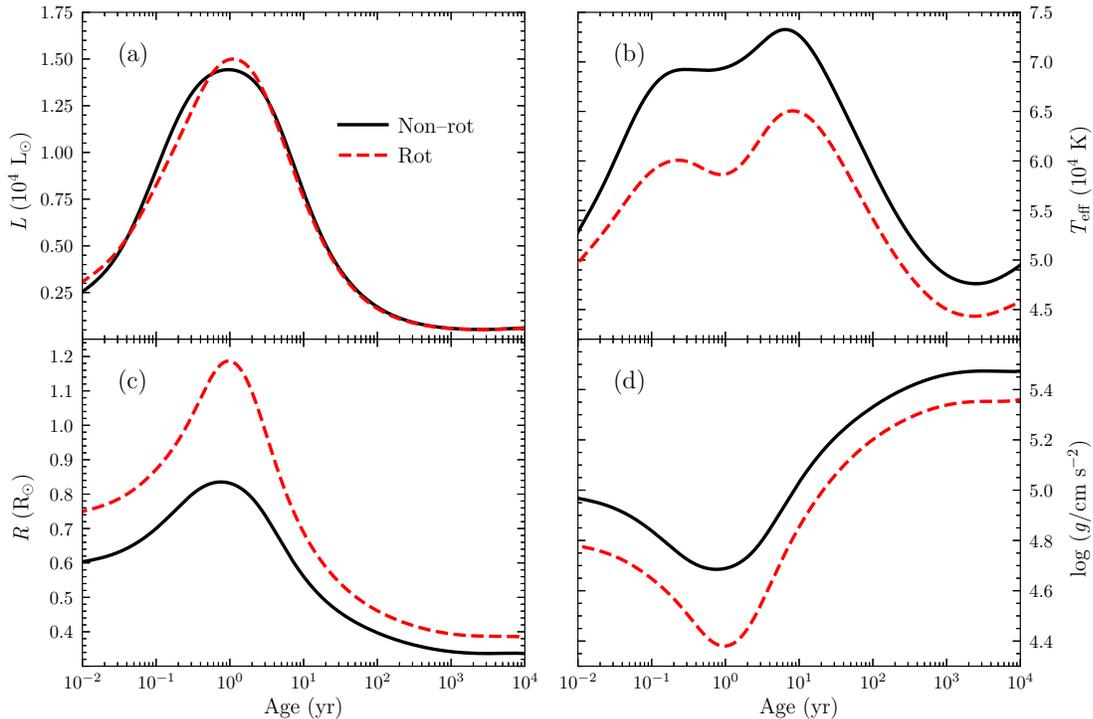}
	\caption{\label{fig:3} Time evolution of the post-impact luminosity, temperature, radius and surface gravity of the rotating surviving He companion model (red dashed line). For comparison, the results of the post-impact evolution of the corresponding non-rotating model are drawn in black solid lines.}
\end{figure*}

\begin{figure*}[t]
	\centering
	\includegraphics[width=1.0\textwidth]{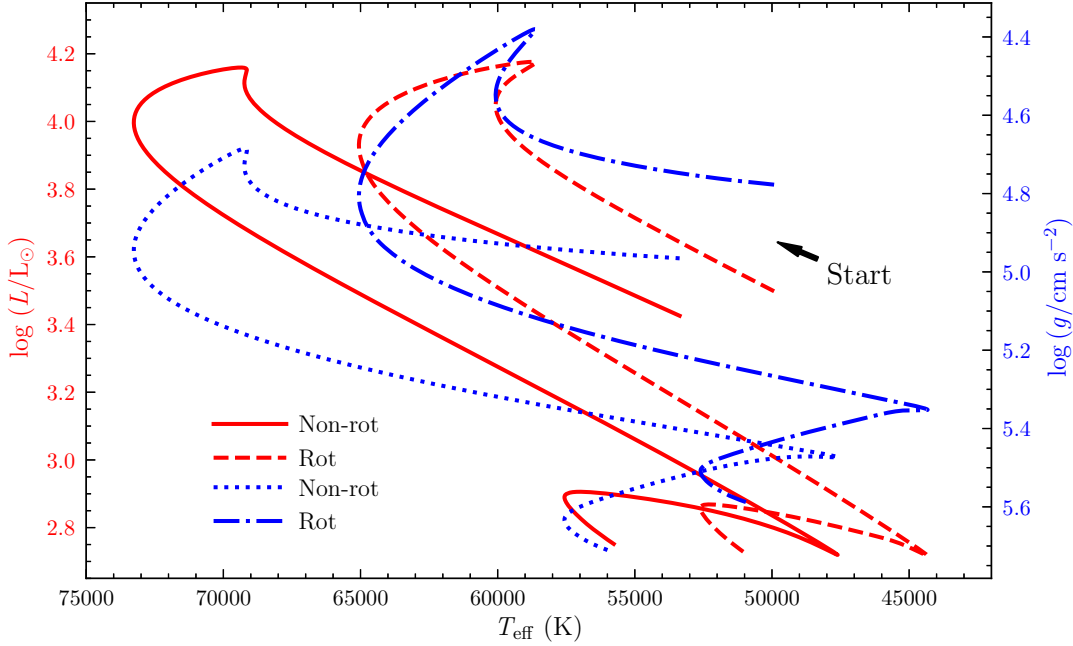}
	\caption{\label{fig:4} Post-impact evolutionary tracks of the rotating companion model in the HR diagram (red dashed lines) and surface gravity vs. temperature diagram (blue dash-dotted lines). For comparison, the results of the corresponding non-rotating model are also drawn in red solid and blue dotted lines.}
\end{figure*}

\subsection{From 3D to 1D}

To present the post-explosion properties of a surviving companion star for their identifications in historical SNRs, we need to follow post-explosion evolution of a surviving companion star for a long timescale up to a few thousand years.  However, it is really difficult to do that by tracing our 3D hydrodynamical simulations for such long timescale because of the very expensive computational cost.  We therefore map the outcome of 3D SPH simulation into the 1D stellar evolution code \textsc{MESA} \citep{Paxton2011ApJS, Paxton2013ApJS, Paxton2015ApJS, Paxton2018ApJS, Paxton2019ApJS} to follow the post-impact evolution of a surviving He-star companion in an SN Iax for a long timescale up to about $10^{5}\,\mathrm{yr}$ until it has relaxed back into thermal equilibrium \citep[see also][]{Liu2021MNRAS,Liu2021b}. 

Whether or not the progenitor binary system could be destroyed after the SN explosion strongly depends on the kick velocity of the bound WD remnant because that the companion star receives a small kick velocity and only a small amount of companion mass is stripped off during the interaction. However, the kick velocity of the bound remnant predicted by current studies is quite uncertain. For instance, \citet{Fink2014MNRAS} obtained a small kick velocity of $\sim36\,\mathrm{km\,s^{-1}}$, but \citet{Jordan2012ApJL} predicted a large kick velocity up to $\sim520\,\mathrm{km\,s^{-1}}$. We refer to \citet[][see their Section.~5.4]{Liu2013ApJb} for a detailed discussion on the fate of the progenitor binary system in this scenario. In this work, we simply assume that the binary system is destroyed after the explosion, and we therefore do not include the bound WD remnant into our 3D impact simulations and the post-impact evolution calculations of surviving He-star companions. The details of the post-explosion evolution with the inclusion of the bound WD remnant should be addressed in future work.

Following the method of \citet[][see their Section.~2]{Liu2021MNRAS}, the angle-averaged radial profiles of internal energy, chemical composition and angular momentum of the surviving He-star companion at the end of our 3D SPH impact simulations are used as inputs of relaxation routines in \textsc{MESA} \citep[][see appendix B]{Paxton2018ApJS} to construct a starting model for its subsequent 1D long-term post-impact evolution \citep[see also][]{Liu2021b,Liu2022ApJ}. Figure~\ref{fig:1} shows a comparison of density distributions of our surviving He-star companion at the pre-explosion phase and at the end of our 3D hydrodynamical simulations. In Figure~\ref{fig:2}, we plot the 1D radial profiles of density, specific internal energy and specific angular momentum of our surviving He-star companion at the pre-explosion phase, the end of 3D impact simulation and the starting phase of the subsequent long-term evolution by \textsc{MESA} calculation.

\section{Results and discussions}
\label{sect:results}

\begin{figure}[t]
	\centering
	\includegraphics[width=1.0\textwidth]{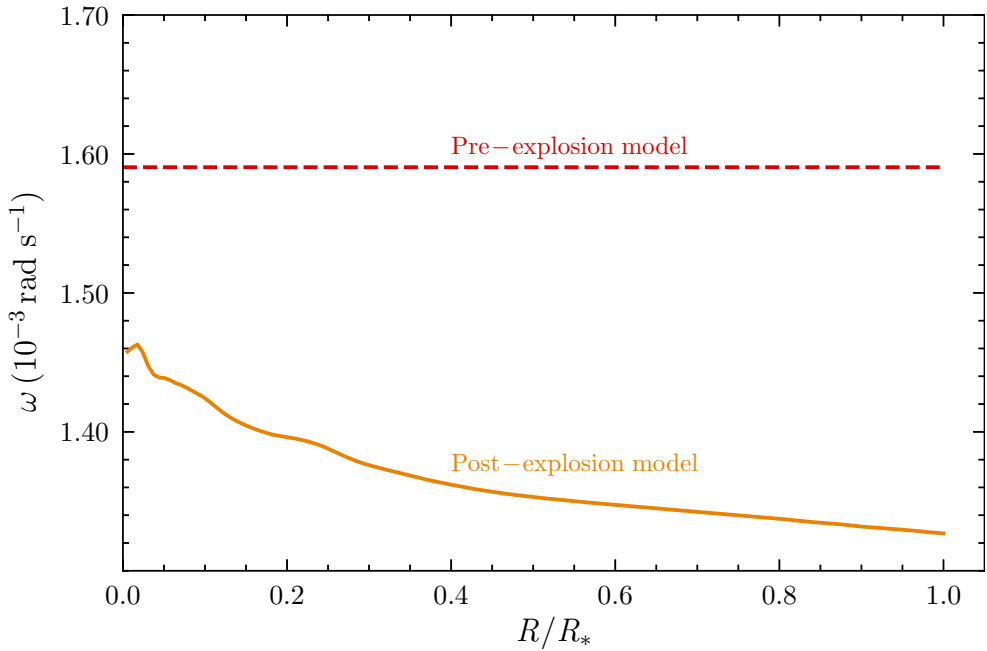}
	\caption{\label{fig:5} Radial profiles of the angular velocity of the pre- (red dashed line) and post- (orange solid line) explosion of the He-star companions. Note that the $x$-axis is normalized to the radius of the He-star companion ($R_{\ast}$).}
\end{figure}

\begin{figure*}[t]
	\centering
	\includegraphics[width=1.0\textwidth]{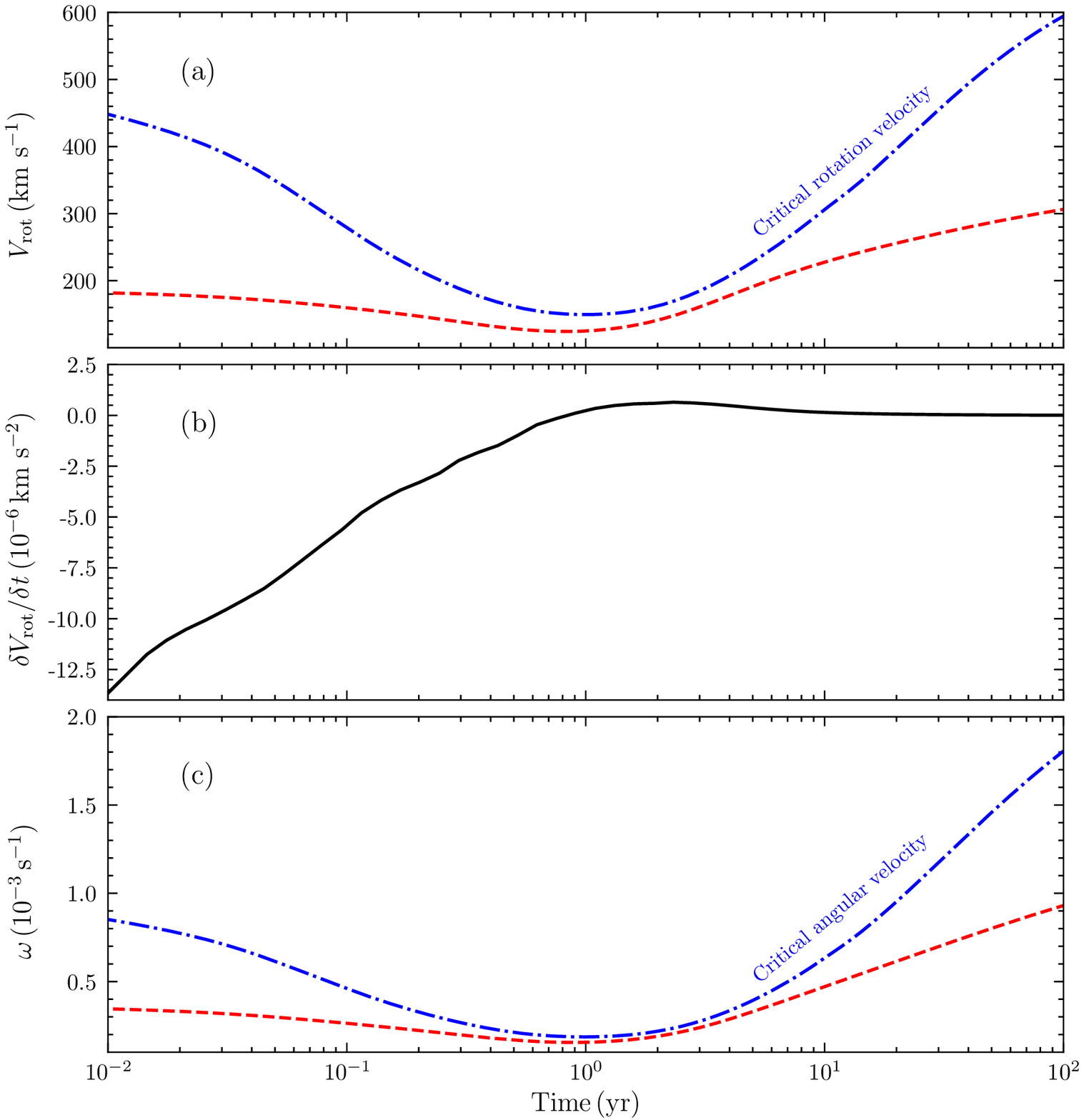}
	\caption{\label{fig:6} Post-impact evolution of the surface rotational speed ($V_{\mathrm{rot}}$, top-panel),  the rate of change of surface rotational speed ($\delta{V}_{\mathrm{rot}}/\delta{t}$, middle-panel) and the surface angular velocity ($\omega$, bottom-panel) of the surviving He-star companion from our MESA calculations. For comparison, the corresponding critical rotational and surface angular velocities are also drawn in dash-dotted lines. The $x$-axis gives the evolutionary time of a surviving He-star companion since the end of our 3D impact simulations ($\sim5\mathord,000\,\mathrm{s}$). For better visibility, the evolution before $10^{-2}\,\mathrm{yr}$ is not shown because no significant difference is observed there.}
\end{figure*}

In this section, we present the results of long-term post-explosion evolution of a surviving He-star companion with 1D \textsc{MESA} calculation, including the post-explosion evolutionary tracks of a star in the Hertzsprung-Russell (HR) diagram and its rotational properties. We also compare the results with those of non-rotating model in our previous study of \citet{ZengInprep}. Our main results are summarized in Table~\ref{Tab:1}.

\subsection{Post-explosion evolutionary tracks}

Based on our 3D impact simulation, we find that the inclusion of the orbital and spin velocities of the progenitor system does not significantly affect the total stripped He masses and the kick velocity received by the star during the ejecta-companion interaction. This is because that the orbital ($\sim430\,\mathrm{km\,s^{-1}}$) and spin ($\sim301\,\mathrm{km\,s^{-1}}$) velocities of the companion star are one order of magnitude lower than the typical velocity of SN ejecta of $\sim7\mathord,000\,\mathrm{km\,s^{-1}}$. We find that about $0.4\%$ of companion mass is removed by the SN explosion, and the star receives a kick velocity of $\sim8\,\mathrm{km\,s^{-1}}$ in both the non-rotating and rotating model.  At the end of 3D SPH impact simulation, the companion star is out of thermal equilibrium because of the mass-stripping and shock heating during the interaction (see Figure~\ref{fig:1}).

In Figure~\ref{fig:3}, we depict the post-explosion luminosity, temperature, photosphere radius and surface gravity of our surviving He-star companion as functions of time. In addition, its evolutionary track in the HR diagram is presented in Figure~\ref{fig:4}. The surviving He-star companion significantly expands after the impact due to the release of the energy deposition by the shock heating during the ejecta-companion interaction. About one year after the explosion, the star reaches a maximum luminosity of $1.5\times10^{4}\,\rm L_{\odot}$ when it expands to a maximum radius of $\rm 1.2\,R_{\odot}$. The star starts to shrink and relax back into its thermal equilibrium as the deposited energy radiates away, and it appears as an O-type hot subdwarf (sdO) star when it has re-established its thermal equilibrium at about a few $1\mathord,000\,\mathrm{yr}$. After that, the star keeps evolving by following a quite similar evolutionary track of a non-impacted He-star with the same mass. This indicates that it would be more difficult to successfully identify the surviving He-star companions after they have relaxed back into thermal equilibrium about a few thousand years after the explosion, because their post-impact evolutionary tracks are almost identical to those of normal He-stars at this late-time phase.  We therefore conclude that the identification of the surviving He-star companions of SNe Iax is more likely to be successful in young nearby SNRs.

\begin{table}[t]
\begin{center}
\caption[]{Results of our simulations.}\label{Tab:1}
 \begin{tabular}{lcccccccc}
  \hline
  \hline\noalign{\smallskip}
   Model & $\Delta M/M_{2}$  & $V_{\rm kick}$ & $E_{\rm inj}$ & $d$ & $\log\,\mathit{L}^{\mathrm{peak}}$ & $\log\,\mathit{T}^{\mathrm{peak}}_{\mathrm{eff}}$ & $\mathit{R}^{\mathrm{peak}}$ & $\mathit{t}^{\mathrm{peak}}$ \\
         & (\%) & ($\mathrm{km\,s^{-1}}$) & ($10^{48}\,\mathrm{erg}$) & ($M_{r}/M_{\ast}$) & ($\mathrm{L_{\odot}}$) & ($\mathrm{K}$) & ($\mathrm{R_{\odot}}$) &  $(\mathrm{yr}$) \\
  \hline\noalign{\smallskip}
   Non-rotating  & $0.4$ & $7.6$ & $3.78$ & $0.995$ & $4.16$ & $4.86$ & $0.84$ & $0.95$ \\ 
   Rotating  & $0.4$ & $8.5$ & $3.89$ & $0.992$ & $4.18$ & $4.81$ & $1.19$ & $1.10$ \\
  \noalign{\smallskip}\hline \\
\end{tabular}
\tablecomments{\textwidth}{$\Delta M$, $V_{\rm kick}$, $E_{\rm inj}$ and $d$ are the amount of stripped He mass, kick velocity, total deposited energy and depth of energy deposition obtained from our 3D impact simulation respectively; $\mathit{L}^{\mathrm{peak}}$, $\log\,\mathit{T}^{\mathrm{peak}}_{\mathrm{eff}}$ and $\mathit{R}^{\mathrm{peak}}$ respectively represent the maximum luminosity, effective temperature and radius of the star during its thermal re-equilibration phase; $\mathit{t}^{\mathrm{peak}}$ gives the time at the maximum luminosity.}
\end{center}
\end{table}

For comparison, the corresponding results of the non-rotating model obtained from our previous work of \citet{ZengInprep} are also drawn in red dashed lines in Figure.~\ref{fig:3}. These differences between rotating and non-rotating model are caused by the amount and depth of energy deposition, and the centrifugal force due to the spin. The depth of energy deposition in the rotating model (which is found to be at $M_{\mathrm{r}}/M_{\ast}\sim0.992$, where $M_{\mathrm{r}}$ is the enclosed mass within a sphere of radius $r$, and $M_{\ast}$ is the total mass of the star) is slightly deeper than that of the non-rotating model (at $M_{\mathrm{r}}/M_{\ast}\sim0.995$), which leads to the non-rotating model taking a slightly shorter timescale to radiate away the deposited energy.  In addition, we find a slightly higher amount of energy deposition (i.e., $3.89\times10^{48}\,\mathrm{erg}$) in the rotating model than that of the non-rotating model (i.e., $3.78\times10^{48}\,\mathrm{erg}$), which may be caused by a slightly bigger cross-sectional area in the rotating model due to its motion. Compared with the non-rotating model, the centrifugal force and a higher energy deposition in the rotating model make its photosphere expand more as what has been observed in Figure~\ref{fig:3}.

\subsection{Post-explosion surface rotational speed}

At the beginning of our 3D impact simulations, the companion star is spherically symmetric (see left-panel of Figure.~\ref{fig:1}), and its initial rotation was set up to be a rigid-body rotation.  As displayed in Figure~\ref{fig:5}, the entire star has the same angular velocity from the center to the surface, and the rotational velocity increases linearly with radius. After the impact, the He-star companion does not rotate as a rigid body (orange solid line in Figure~\ref{fig:5}), and some differential rotational features appears. If the He-star companion has an initial magnetic field, the differential rotational features might enlarge the magnetic field \citep{Spruit2002AAP}. We find that about $2\%$ of initial angular momentum of the companion star is lost because that about $0.4\%$ of its mass is removed during the interaction at the end of our impact simulation. Meanwhile, the companion radius increases by a factor of $2.5$ to reach $\sim0.68\,\mathrm{R_{\odot}}$ due to the shock heating. This leads to the surface rotational velocity of the star dropping to $\sim180\,\mathrm{km\,s^{-1}}$ at the end of our impact simulation from its pre-explosion value of $\sim301\,\mathrm{km\,s^{-1}}$.

In Figure~\ref{fig:6}, we show the post-impact evolution of the surface rotational speed and angular velocity of the surviving He-star companion from our \textsc{MESA} calculation.  As the star expands, its surface rotation keeps decreasing and reaches the minimum value of $\sim100\,\mathrm{km\,s^{-1}}$ (which is about $1/3$ of its pre-impact value) when the star expands to a maximum radius at a few years after the impact. After that, the star starts to shrink as the deposited energy is radiated away, leading to the star spinning up again. About $10^{2}\,\mathrm{yr}$ after the impact, it becomes a fast rotator again and has a surface rotational speed of about $300\,\mathrm{km\,s^{-1}}$, and the rate of change in surface rotational speed ($\delta{V}_{\mathrm{rot}}/\delta{t}$) is close to zero (see the top and middle panels of Figure.~\ref{fig:6}). We therefore conclude that this rotation-switching feature within a few years may provide a useful way to identify the surviving He-star companions of SNe Iax in future observations.

The He-star companions are expected to have the rotational speeds of $140$--$380\,\mathrm{km\,s^{-1}}$ at the moment of SN explosion based on binary population synthesis calculations for the WD~+~He-star channel of SNe Ia \citep{Wang2009AAP}, which is slower than the typical velocity of SN ejecta of $\sim7\mathord,000\,\mathrm{km\,s^{-1}}$ by one order of magnitude. Therefore, we do not expect that the inclusion of the rotational speed of a companion star would significantly affect the results of the ejecta-companion interaction such as the total stripped He mass, the kick velocity, and the amount and depth of energy deposition, compared with those of the non-rotating model. In addition, the fastest spinning companion star in the WD~+~He-star channel has a rotational velocity of $380\,\mathrm{km\,s^{-1}}$, which is not significantly faster than our model ($300\,\mathrm{km\,s^{-1}}$) in this work. We therefore do not expect that the post-impact properties of our rotating model would change significantly if it spins with a bit higher velocity of $380\,\mathrm{km\,s^{-1}}$. For the slowest spinning model, we expect that it would have similar post-impact properties with our non-rotating model. However, we expect that various spin speeds would cause different centrifugal forces and thus lead to the star having a slightly larger/smaller radius compared with that of the rotating model in this work. \citet{Liu2013AAP} find that the post-explosion rotational speeds are scaled linearly with the pre-explosion rotational speeds for surviving main-sequence stars. If we simply assume that such a relation still holds for surviving He-star companions in this work, we could roughly estimate that the rotational velocities of the surviving He-star companions could drop to $47$--$127\,\mathrm{km\,s^{-1}}$ after the impact.

\section{Summary and Conclusion}
\label{sect:summary}
In this work, we have performed a 3D hydrodynamical simulation of the interaction of SN ejecta with a He-star companion with the SPH code \textsc{Stellar GADGET} \citep{Pakmor2012MNRAS} by directly adopting the He-star companion model and explosion  model used by \citet{Zeng2020ApJ}. However, the rotation of the companion star and orbital motion of the binary system are also taken into account in the 3D impact simulation of this work. We further follow the long-term post-impact evolution of the surviving companion star by using 1D stellar evolution code \textsc{MESA} \citep{Paxton2011ApJS, Paxton2013ApJS, Paxton2015ApJS, Paxton2018ApJS, Paxton2019ApJS}. We aim to focus on exploring the post-impact rotation evolution of the surviving He-star companions of SNe Iax. Our results and conclusions are summarized as follows.

\vspace{-\topsep}
\begin{itemize}

\item[\textbullet] We find that about $0.4\%$ of companion masses are removed by SN Ia impact during the ejecta-companion interaction, and the He-star companion receives a kick velocity of $\sim8\,\mathrm{km\,s^{-1}}$. These results are comparable to those of the non-rotating model of \citet{Zeng2020ApJ}.

\item[\textbullet] We find that the depth of energy deposition due to the shock heating during the interaction in the rotating model is at about $M_{\mathrm{r}}/M_{\ast}\sim0.992$, which is slightly deeper than that of the non-rotating model of $M_{\mathrm{r}}/M_{\ast}\sim0.995$. In addition, the amount of energy deposition in the rotating model (i.e., $3.89\times10^{48}\,\mathrm{erg}$) is higher than that of the non-rotating model (i.e., $3.78\times10^{48}\,\mathrm{erg}$). A higher amount of energy deposition and the centrifugal force in the rotating model make its photosphere expand more compared with the non-rotating model (Figure~\ref{fig:3}).

\item[\textbullet] About $2\%$ of initial angular momentum of the companion star is lost because that about $0.4\%$ of its mass is removed by the SN Ia impact. In addition, the companion star significantly puffs up due to the shock heating. As a result, the surface rotational velocity of the companion star drops to $\sim180\,\mathrm{km\,s^{-1}}$ at the end of our impact simulation from its pre-explosion value of $\sim301\,\mathrm{km\,s^{-1}}$.

\item[\textbullet] The surface rotational velocity of the surviving He-star companion keeps decreasing as it expands,  and it reaches the minimum value of $\sim100\,\mathrm{km\,s^{-1}}$ when the star expands to a maximum radius at about a few years after the impact. Subsequently, the star starts to shrink as the deposited energy is radiated away, leading to the surviving He-star companion spinning up again. Our results suggest that the rotation of the surviving He-star companions of SNe Iax could significantly drop, although they were originally fast-rotating at the moment of SN Ia explosion.

\item[\textbullet] We find that the surviving He-star companions of SNe Iax experience the spin-down and spin-up phases within a few years after the explosion. This peculiar rotation-switching feature would be useful for the identification of surviving He-star companions of SNe Iax in future observations.

\end{itemize}

\normalem
\begin{acknowledgements}

We thank the anonymous referee for his/her constructive comments that helped to improve this paper. This work is supported by the National Key R\&D Program of China (Nos. 2021YFA1600400, 2021YFA1600401), the National Natural Science Foundation of China (NSFC, Grant Nos.~11873016, 11973080, and 11733008), the Chinese Academy of Sciences, and Yunnan Province (Nos.~12090040, 12090043, 202001AW070007, 2019HA012, and 2017HC018). X.M.\ acknowledges support from the Yunnan Ten Thousand Talents Plan--Young \& Elite Talents Project, and the CAS `Light of West China' Program. The authors gratefully acknowledge the ``PHOENIX Supercomputing Platform'' jointly operated by the Binary Population Synthesis Group and the Stellar Astrophysics Group at Yunnan Observatories, Chinese Academy of Sciences. This work made use of the Heidelberg Supernova Model Archive \citep[HESMA,][see \url{https://hesma.h-its.org}]{Kromer2017MmSAI}.

\end{acknowledgements}
  

\end{document}